# Simulating geomagnetic reversals through 2D Ising systems

Jorge O. O. Franco[1,2], Vitor H. A. Dias[1,2] and Andrés R. R. Papa[1,2,*]

[1]Observatório Nacional, Rua General José Cristino 77, São Cristóvão,
Rio de Janeiro, 20921-400 RJ, BRASIL
[2]Instituto de Física, Universidade do Estado do Rio de Janeiro, Rua São Francisco Xavier 524, Maracanã,
Rio de Janeiro, 20550-900 RJ, BRASIL

**Abstract**

In this work 2D Ising systems were used to simulate the reversals of the Earth's magnetic field. Each spin was supposed to be one ring current in the Earth dynamo and the magnetization was supposed to be proportional to the field intensity. Given the relative success of some physical few-discs modeling of this system all the simulations were implemented in small system. The temperature T was used as a tuning parameter. It plays the role of external perturbations. Power laws were obtained for the distribution of times between reversals. When the system size was increased the exponent of the power law asymptotically tended towards values very near −1.5, generally accepted as the right value for this phenomenon. Depending on the proximity of T and $T_C$ the average duration of reversal periods changes. In this way it is possible to establish a parallel between the model and more or less well defined periods of the reversal record. Some possible trends for future works are advanced.

___________

* E-mail address: papa@on.br

# 1. Introduction

The Ising model is some kind of paradigm among magnetic models for the matter because is the simplest one that presents a phase transition at temperatures different from zero (in D>1, where D is the dimension). But Ising models have found applications in areas away from solid-state physics, for example, in immunology [1] and stock market theory [2], among many others. In this work we use the 2D Ising model to simulate the Earth's dynamo, particularly the reversals of the Earth's magnetic field.

In absence of external field the Hamiltonian $H$ for the Ising model is:

$$H = -\sum_{i,j} J_{ij} s_i s_j \qquad (1)$$

where the sum runs over all neighbor pairs, $s_i$ and $s_j$ are spins and $J_{ij}$ is the interaction constant between spins $i$ and $j$. The critical temperature of the Ising system for 2D is $T_C = 2.269$ in units of $J_{ij}/k_B$, where $k_B$ is the Boltzmann constant. By definition, each spin $s_i$ can assume only two values or orientations: 1 and –1.

The magnetization $M$ for the model is defined as:

$$M = \frac{\sum_i s_i}{N} \qquad (2)$$

where the sum runs over all the spins and N is the total number of spins. It can takes values between –1 and 1. While in our model each spin represents a current ring in the dynamo, the magnetization represents the magnetic field intensity.

Geomagnetic reversals, periods during which the Earth's magnetic field swap hemispheres, have been documented during the last four decades. The duration of geomagnetic reversals presents power law distribution functions, which can be the signature of self-organized criticality as the mechanism of their source [3,4]. The sequence of reversals been considered as first returns times. In Figure 1 we present the sequence of geomagnetic reversals from 80 Ma to our days. The data was obtained from the more recent and complete record that we have found [5]. However, the behavior is much more complex than that, it is necessary a finite time for the magnetic field changing its sign, the change is not instantaneous. Actually, the magnetic field takes continuous values, and not only, let us say, a maximum positive value and a minimum negative value (as most part of the time in Ising systems below the critical temperature). There are some facts that can be argued against the possibility of self-organized criticality. In particular, that there are clustered periods of high activity (for example, from 40 My ago to now and during the period 165–120 My), periods of low activity (between 80 and 40 My ago) and periods of almost null activity (from 120 to 80 My). Self-organized criticality implies that there are neither characteristic lengths nor characteristic times [6]. Activity occurs all the time in all the possible length and time scales. The unique relevant length scale is the size of the system (intimately related to the duration of the more lasting events). There are other mechanisms capable of producing power laws (for example, superposition of some distributions [7] and non-extensive versions of statistical mechanics [8]) but we will not extend on them upon here.

Many efforts have been made to enlarge the size of simulated Ising models when trying to represent systems of physical or biological nature. The reason is simple: the numbers of elements in condensed matter physics and biological systems are usually huge, around $10^{23}$ and $10^{14}$, respectively. Physical models for the Earth's magnetic field are mostly based in a small number of adequately connected dynamos. So, using small 2D Ising systems to simulate their interaction and behavior must sound a natural choice.

Ising systems have already been used to simulate reversals of the Earth magnetic field [9]. Our work differs from it in that while they used relatively large systems though the Q2R automata we have simulated a regular Metropolis updating algorithm on small systems around the critical temperature. Our results also essentially differ from them.

## 2. Simulations and Discussion

We have performed simulations on small 2D square Ising systems at temperatures near $T_C$ and studied some of the relevant characteristics of the distribution in time of magnetization reversals (fluctuations between states of positive and negative magnetization when $T<T_C$). All the simulations run on small Ising systems. We have applied periodic boundary conditions [10].

In Figure 2 we show typical magnetization dependences on time for temperatures slightly below $T_C$. They qualitatively resemble the results presented in figures 3 and 5 of reference [9]. However, they are quantitatively different from them (see below). Figure 3 shows the result of eliminating all the fluctuations in Figure 2. We have only marked the jumps through the value zero and assigned a value 1 to all the positive values and −1 for the negative ones. Note the pictorial and qualitative similarity between figures 1 and 3. In particular, note that Figure 3 also presents clustered periods of greater activity (between ~6500 and 10000 MCS/s) and of lower activity (between ~4000 and 6500 MCS/s, for example).

Figure 4 shows the distribution of intervals between consecutive reversals (T=2.26) for different system sizes. The bold straight line is a guide for the eye in all cases. It is a general feature that for small periods the distribution follows a power law:

$$f(t) = ct^d \qquad (3)$$

where *f(t)* is the frequency distribution of periods between consecutive reversals, *c* is a proportionality constant and *d* is the exponent of the power-law. For larger periods the distribution is an exponential. The exponential dependence can easily be confirmed be representing the data in a semi-log graph (not shown). However its apparition presents a size dependence, for 5x5 systems it occupies almost all the histogram interval while for systems of size 50x50 it is almost inexistent. The results for the slope value range from –2.0 (for 5x5 systems) to –1.55 (for 50x50 systems). These results are far from the –0.5 value obtained by Seki and Ito [9]. However, as can be seen, the larger the system the closer to the most accepted value (–1.5) for this slope. There are some recent results [4] suggesting that a value –1.42 might be more reliable.

To complete this part of our research we have also performed a no pretentious study on sizes effects. For this we have simulated system sizes 5x5, 10x10, 20x20, 30x30, 40x40 and 50x50 for the same temperature T = 2.26 (very near $T_C$ to avoid huge simulation times for the largest systems). The results are shown in Figure 5. The data is satisfactorily fitted by an exponential function.

We show in Figure 6 the distribution of magnetization values (T=2.26) for 5x5 and 50x50 systems. For the smallest system that we have simulated the maxima occur at the extreme values M = ±1. For larger systems they occur at intermediate values. This can be partially explained by fluctuations. While for small system fluctuations can drive the system to absolute values of magnetization near 1, and there are not a counterpart, for larger system, a fluctuation to an absolute value near 1 in a part of the system will be accompanied by a value far from those values in other pieces of the system. We have not risked fit the distribution by any function because our model introduced and artificial cut-off for the field intensity (maximum absolute magnetization 1). Contrary to the record of reversals (from 160 My up to nowadays) the record of magnetic field intensities exists just for a few thousand years. The distribution of magnetic field intensities for actual reversals follows a function between a normal and a log-normal distribution. However, the number of experimental points is small and there will certainly be changes in these facts when new measurements become available. There are two points to be highlighted: as in our simulations the distribution of intensities for real measurements is almost perfectly symmetric and there is a valley (less frequent) for values near zero intensity. The symmetry is not observed in the work by Seki and Ito [9].

Figure 7 shows the dependence of the average duration of time intervals between consecutive reversals on temperature (always for temperatures near the critical point). These results can help us to understand the way in which external factors affect the reversal regime. For more intense perturbations (high temperatures) the reversals are more frequent. For less intense perturbations (low temperatures) there is a lower number of reversals for time unit.

In our simulations the temperature plays the role of a tuning parameter. For temperatures well above $T_C$ we observe only fluctuations around M = 0, for temperatures well below $T_C$ we observe a state of constant magnetization values 1 or –1 with seldom flips to –1 or 1 values, respectively. We have studied the dependence of the power-law slope on temperature for temperatures T around $T_C$ (Figure 8). Note that for temperatures near $T_C$ or above the slope remains almost constant and in a narrow interval (from –1.75 to –1.55). The model is robust (in this particular) to temperature variations. The point at T = 1.23 most be regarded with some suspicion. It contradicts, at least intuitively, what we could expect: for low temperatures reversals are rare. This means that small and large intervals between consecutive reversals must have similar probabilities consequently diminishing the slope of the distribution function. We believe that for that temperature our simulations did not run for long enough times.

Finally, Figure 9 presents the mean interval between consecutive reversals as a function of the system size. That dependence gives some qualitative insight on which is the volume of Earth's interior that effectively carries the currents that produce the magnetic field. For larger systems reversals are less frequent than for small systems.

## 3. Conclusions

We have simulated small 2D Ising systems to mimic the time dependence of geomagnetic reversals. There are several things to be learnt from our simulations. The almost perfect symmetry of magnetization distributions is the fingerprint that our simulations were running for a long enough time. Any future work along these lines should reproduce our results. The model seems to be robust to temperature variations (in the slope of the distribution for intervals between consecutive reversals). There is a dependence of the mean interval between reversals in both temperature and system size. These facts make difficult to distinguish which of them can be the cause of the clustering observed in actual reversals. The necessity of more detailed models including dissipation is apparent. Some works along these lines are currently running and will be published elsewhere.

## Acknowledgements

The authors sincerely acknowledge partial financial support from FAPERJ (Rio de Janeiro Founding Agency) and CNPq (Brazilian Founding Agency).

**Figure Captions**

Figure 1.- Representation of geomagnetic reversals from around 80 My ago to our days. We arbitrarily have assumed −1 as the current polarization. We have not plotted the period from 80 My to 165 My (for which also exist record) for the sake of clarity. There was a period with no reversals from ~ 80 to 120 My (the beginning appears in the plot). The period from 120 My to 165 My was very similar to the period from 0 to 40 My (in number of reversals and in the average duration of reversals). The data was obtained from Cande and Kent [5].

Figure 2.- Time dependence of the magnetization for temperatures slightly below $T_C$ (T = 2.26). a) A sample of 1000 MCS/s pertaining to a long run. b) A longer sample (up to 10000 MCS/s). Both cases correspond to a 5x5 system size.

Figure 3.- A binary version of the dependence in Figure 2b. It was assigned a value 1 to all values > 0 and a value of −1 for all values < 0. Note the similarity with Figure 1 where actual reversals are presented.

Figure 4.- Distribution of periods between consecutive reversals for our model. The system sizes are 5x5, 10x10 and 50x50 for a), b) and c), respectively. In all cases the bold straight line is a guide for the eye. Slopes values are a) −2.0, b) −1.87 and c) −1.55. Runs were done at a fixed T=2.26.

Figure 5.- Size dependence for the slope of the distribution of periods between consecutive reversals in our model. The bold curve is an exponential fit to data. It has the form y=$y_0$ + A.exp(-x/$\tau$), where $y_0$= −1.559 ± 0.014, A = −0.457 ± 0.023 and $\tau$ = 326 ± 49.

Figure 6.- Distributions of magnetization values. For 5x5 systems (squares) the maxima appear at the extremes $|M|$=1. For larger systems, and in particular for 50x50 systems (circles), maxima appear at values of magnetization $|M| \neq 1$.

Figure 7.- Temperature dependence of the mean time interval between consecutive reversals. For temperatures below 2.18 there were no reversals (for simulations up to $10^6$ MCS/s). There seems to be an asymptote.

Figure 8.- Temperature dependence of the power-law exponent. A system size 50x50 was used in all cases.

Figure 9.- Size dependence of the mean time interval between consecutive reversals. The same temperature T = 2.26 was used in all cases.

Figure 1

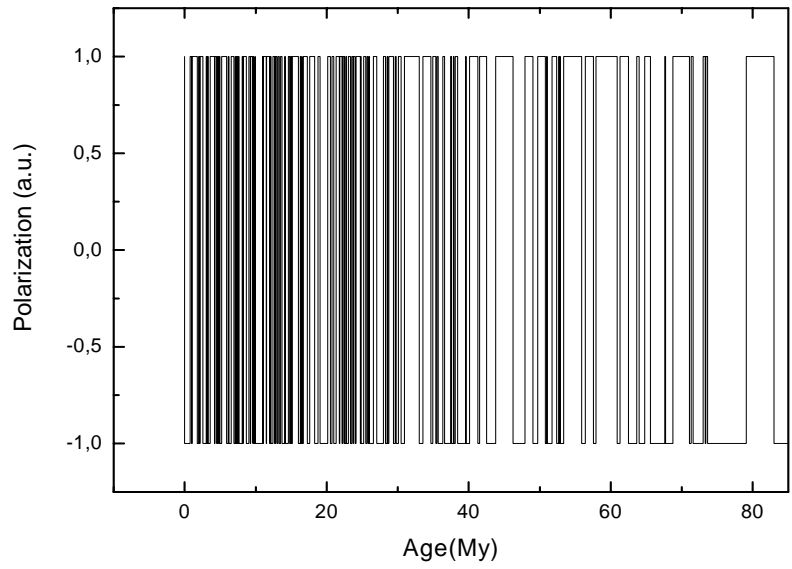

Figure 2

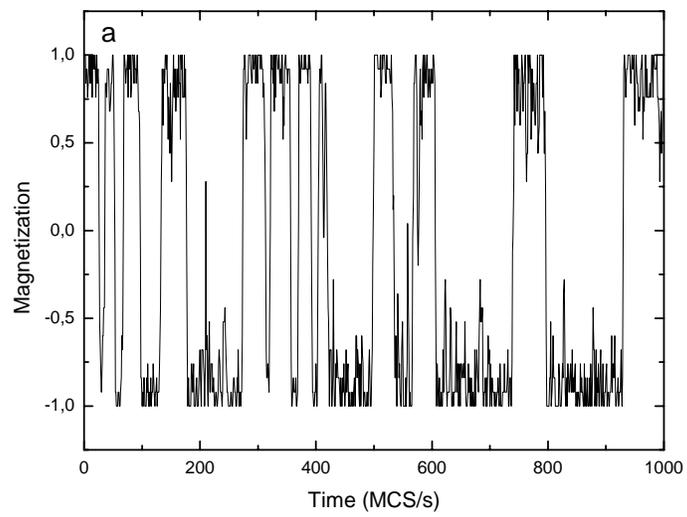

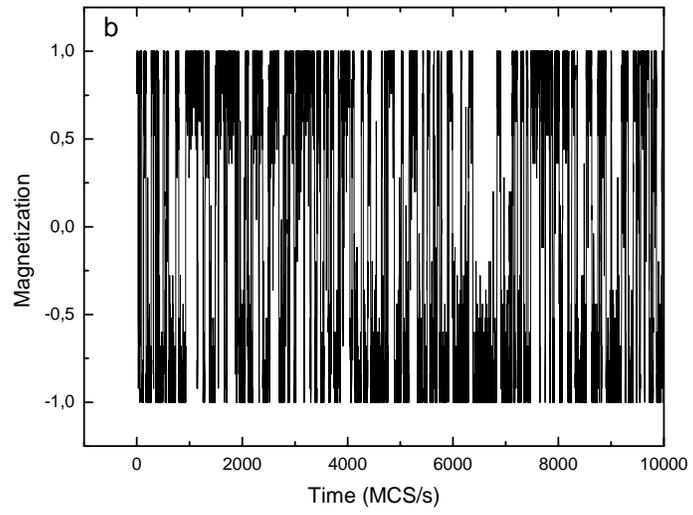

Figure 3

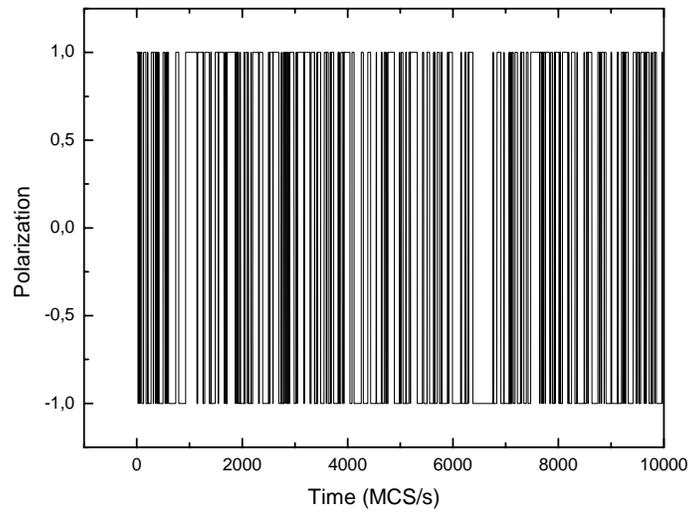

Figure 4

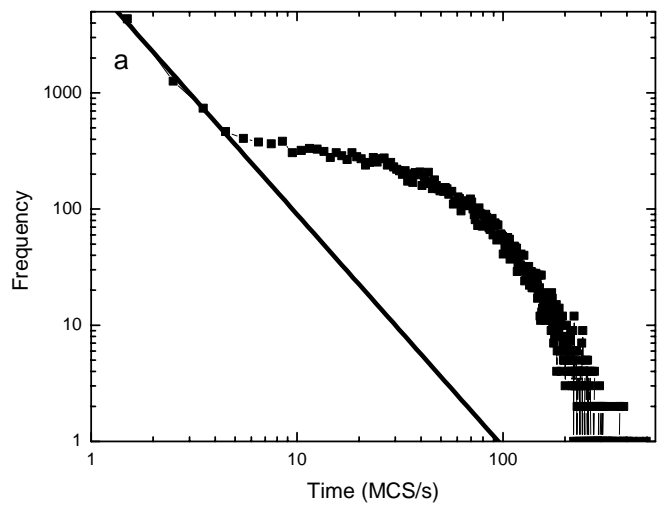

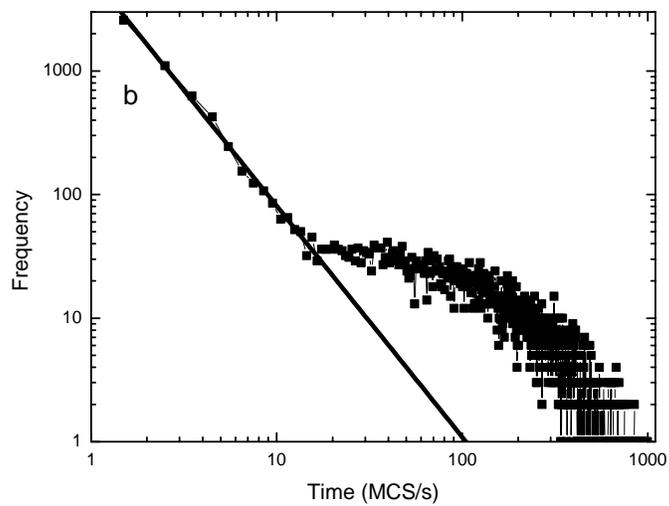

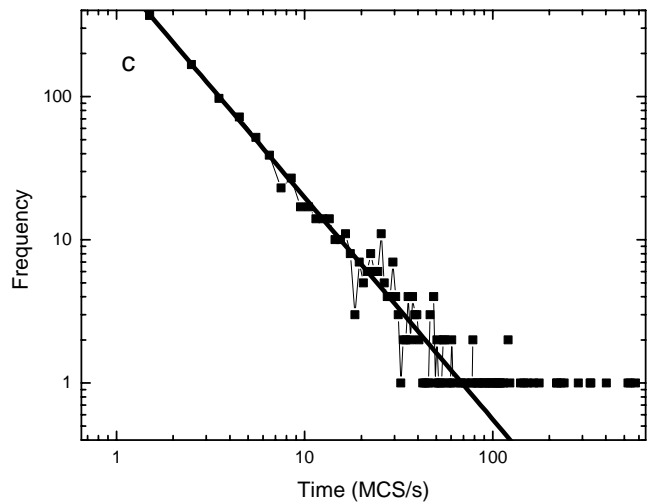

Figure 5

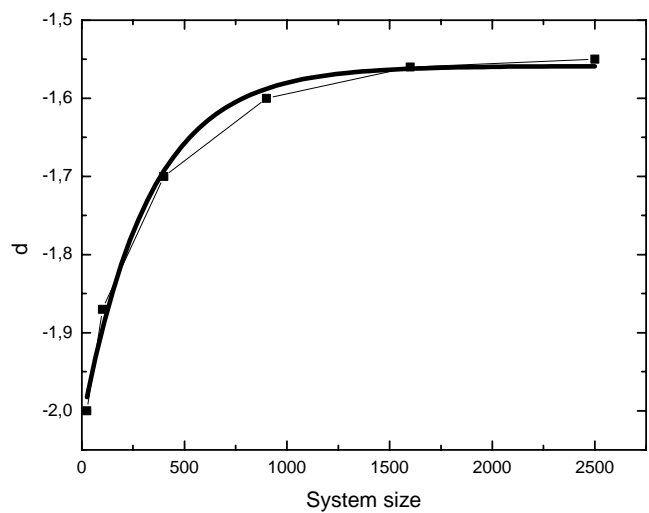

Figure 6

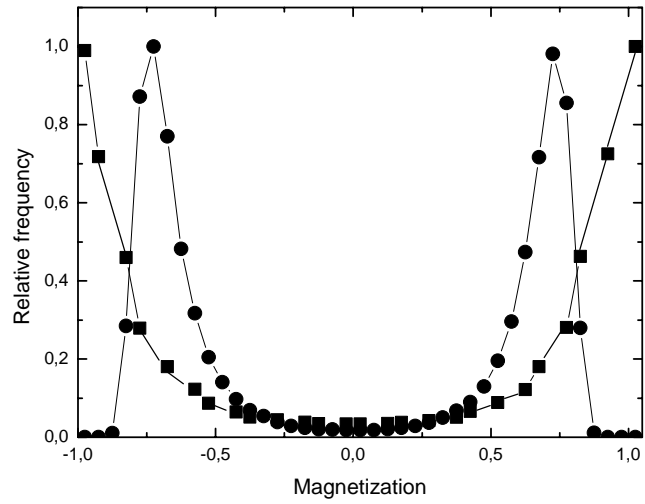

Figure 7

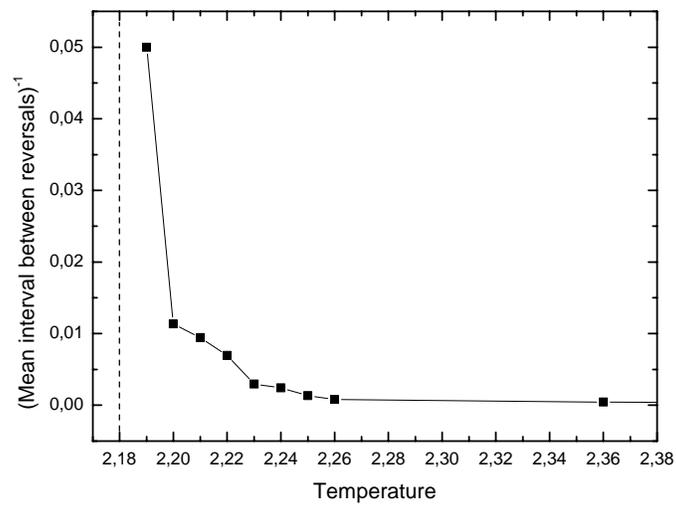

Figure 8

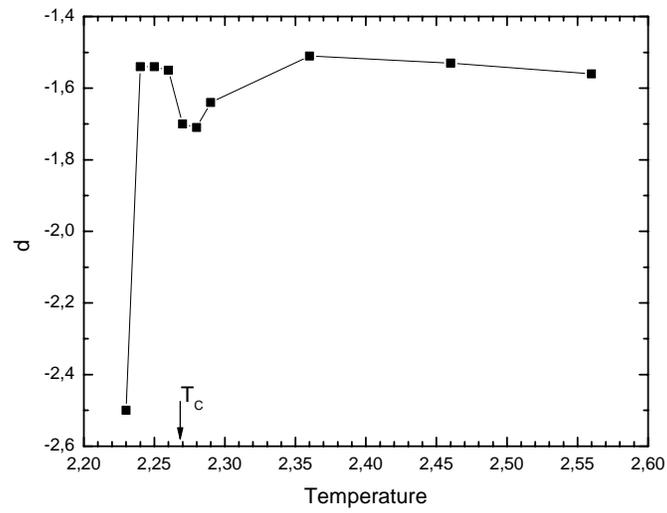

Figure 9

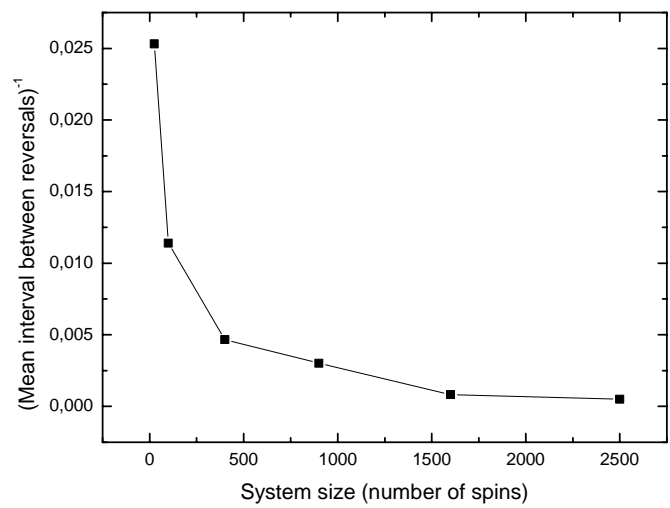